\documentclass[5p]{elsarticle}

\usepackage{amsfonts}
\usepackage{amsmath}
\usepackage{amssymb}
\usepackage{bbold}
\usepackage{bm}
\usepackage{cellspace}
\usepackage{color}
\usepackage{cuted}
\usepackage{enumerate}
\usepackage{graphicx}
\usepackage{physics} 
\usepackage[figuresright]{rotating}
\usepackage{siunitx}
\usepackage[caption=false]{subfig} 
\usepackage{simpler-wick}
\usepackage{xfrac}
\usepackage[pdfencoding=auto, pdfpagelabels]{hyperref}
\hypersetup{breaklinks=true,colorlinks=true,linkcolor=blue,citecolor=blue,
filecolor=magenta,urlcolor=blue}

\graphicspath{{momentum_dist_figures/}} 

\usepackage{simpler-wick}

\newcommand{\Flo}{F^{\rm lo}}
\newcommand{\Fhi}{F^{\rm hi}}

\newcommand{\kvec}[0]{\mathbf{k}}
\newcommand{\kpvec}[0]{\mathbf{k'}}

\newcommand{\Kvec}[0]{\mathbf{K}}

\newcommand{\qvec}[0]{\mathbf{q}}

\newcommand{\fmi}{\ensuremath{\mbox{fm}^{-1}}}

\newcommand{\helium}{\ensuremath{^{4}\text{He}}}
\newcommand{\carbon}{\ensuremath{^{12}\text{C}}}
\newcommand{\oxygen}{\ensuremath{^{16}\text{O}}}
\newcommand{\calcium}{\ensuremath{^{40}\text{Ca}}}

\newcommand{\NNNNLO}{N$^4$LO}

\newcommand{\akdag}[0]{a^{\dagger}}
\newcommand{\ak}[0]{a^{\phantom{\dagger}}}

\newcommand{\akfull}[3]{\ak_{{#1}{#2}{#3}}}

\newcommand{\wcA}[2]{\wick{\c a^{\dagger}_{{#1}} \c a_{{#2}}}}

\newcommand{\melr}[3]{(#1|#2|#3)}

\newcommand{\sg}[1]{\sigma_{{#1}}}
\newcommand{\sgn}[0]{\sigma}
\newcommand{\ta}[1]{\tau_{{#1}}}

\newcommand{\nhat}{\widehat{n}}
\newcommand{\nhathi}{\nhat_{\infty}}  
\newcommand{\nhatlo}{\nhat_{\lambda}}  

\newcommand{\Uhat}[0]{\widehat U}

\newcommand{\Uhatlam}[0]{\Uhat_{\lambda}^{\phantom{\dagger}}}
\newcommand{\Uhatlamdag}[0]{\Uhat_{\lambda}^{\dagger}}
\newcommand{\delUlam}[0]{\delta U_{\lambda}}
\newcommand{\delUdaglam}[0]{\delta U^{\dagger}_{\lambda}}

\newcommand{\Ihat}[0]{\widehat I}

\begin{document}

\title{High-resolution momentum distributions from low-resolution wave functions}

\author[1]{A.~J.~Tropiano\corref{cor1}}
\ead{atropiano@anl.gov}

\author[2]{S.~K.~Bogner}
\ead{bogner@frib.msu.edu}

\author[3]{R.~J.~Furnstahl}
\ead{furnstahl.1@osu.edu}

\author[3]{M.~A.~Hisham}
\ead{hisham.3@buckeyemail.osu.edu}

\author[1,4]{A.~Lovato}
\ead{lovato@alcf.anl.gov}

\author[1]{R.~B.~Wiringa}
\ead{wiringa@anl.gov}

\affiliation[1]{organization={Physics Division},
addressline={Argonne National Laboratory},
city={Argonne, IL},
postcode={60439},
country={USA}}

\affiliation[2]{organization={Facility for Rare Isotope Beams and Department of Physics and Astronomy},
addressline={Michigan State University},
city={East Lansing, MI},
postcode={48824},
country={USA}}

\affiliation[3]{organization={Department of Physics},
addressline={Ohio State University},
city={Columbus, OH},
postcode={43210},
country={USA}}

\affiliation[4]{organization={Computational Science Division},
addressline={Argonne National Laboratory},
city={Argonne, IL},
postcode={60439},
country={USA}}

\cortext[cor1]{Corresponding author}

\date{\today}

\begin{abstract}
Nucleon momentum distributions calculated with a common one-body operator vary with the resolution scale (and scheme) of the Hamiltonian used.
For high-resolution potentials such as Argonne $v_{18}$ (AV18) there is a high-momentum tail, reflecting short-range correlations in the nuclear wave function, which is reduced or absent for softer, lower-resolution interactions.
We explore if the similarity renormalization group (SRG) can be used to quantitatively reproduce the high-resolution distributions from variational Monte Carlo at all momenta using SRG-evolved operators and empirically fit single-particle orbitals rather than a full RG evolution of many-body wave functions.
The goal of this approach is to enable calculations of high-resolution distributions for a wider range of nuclei as well as for other interactions, and provides connections to phenomenological analyses of experiments.
\end{abstract}

\maketitle

\newpage

\section{Introduction}
\label{sec:introduction}

Nuclear experiments often seek to isolate proc\-ess-in\-de\-pendent quantities, which are expressed theoretically as matrix elements of well-defined operators.
For some kinematic regimes these quantities are quark and gluon parton distributions; the analogs for low-energy nuclear physics include nucleon momentum distributions.
For both we need robust factorization of reaction and structure in the analysis.
This separation induces a scale (and scheme) dependence to the distributions but quantities at different scales are naturally related using renormalization group (RG) transformations~\cite{Tropiano:2021qgf}. 
In this paper we use the similarity RG (or SRG)~\cite{Bogner:2006pc,Bogner:2009bt,Furnstahl:2013oba,Hergert:2016iju} to compare single-nucleon momentum distributions calculated at high resolution, which entails matrix elements of a one-body operator in many-body wave functions that include short-range correlations (SRCs), to distributions using SRG-evolved operators evaluated using simple wave functions appropriately matched.
Such an approach would enable comparisons between results from different nuclear Hamiltonians and from phenomenogical analyses of experiments, and extends the range of nuclei for which high-resolution momentum distributions can be calculated.

A high-resolution Hamiltonian is one that couples momentum modes well above the Fermi momentum into low-energy states, in particular inducing SRCs in many-body wave functions.
A prototypical example is the Argonne $v_{18}$ (AV18) nucleon-nucleon (NN) interaction~\cite{Wiringa:1994wb}, which when supplemented with appropriate three-body forces~\cite{Carlson:1983kq,Pieper:2001ap}, has a long history of phenomenological successes~\cite{Carlson:2014vla,Lonardoni:2017egu}.
In conjunction with the generalized contact formalism (GCF)~\cite{Weiss:2015mba,Weiss:2016obx,Weiss:2018tbu,Weiss:2021zyb}, these successes include the reproduction of SRC phe\-nom\-e\-nol\-ogy from recent experiments that kinematically isolate the effects of short-distance physics~\cite{Hen:2016kwk,Korover:2014dma,Hen:2014nza,Duer:2018sby,Duer:2018sxh,Schmookler:2019nvf,Cruz-Torres:2020uke,Schmidt:2020kcl,CLAS:2020rue}.
But this phenomenology was also shown in Ref.~\cite{Tropiano:2021qgf} to be reproduced with simple calculations at low resolution after RG evolution, which shifts the SRCs from wave functions to operators.
(For a simple pedagogical model of these shifts in the context of field theory transformations, see Ref.~\cite{Furnstahl:2001xq}.)
We build on these results to test whether this approach can work quantitatively to reproduce high-reso\-lu\-tion momentum distributions across all momenta without sacrificing the simplicity.

In principle we can exactly reproduce high-resolution reaction calculations by consistently SRG-evolving both the operators and wave functions (or the Hamiltonian used to generate them).
SRG transformations are unitary, thus preserving observables at all scales.
However, treating both structure and reaction parts at the same SRG scale and scheme is required for consistent prediction of observables when comparing different theoretical approaches.
The SRG resolution scale is associated with the flow parameter $\lambda$, which roughly corresponds to the maximum-momentum components in low-energy wave functions of the transformed Hamiltonian.
Note that the RG resolution scale $\lambda$ is not to be confused with the experimental resolution, which is set by the kinematics of the experiment.

The low-SRG-resolution approach naturally describes factorization of mean-field nuclear structure and two-body high-momentum physics associated with SRC pairs, as the SRG transformations themselves factorize when there is a large scale separation in momentum~\cite{Bogner:2012zm,Anderson:2010aq,Tropiano:2020zwb}.
A full SRG evolution would involve tracking induced many-body forces and reaction operators beyond the two-body level, and accurately treating the long-range correlation structure of the many-body wave function.
We seek to incorporate all of the \emph{important} details of a low-SRG resolution wave function by using phenomenological Woods-Saxon orbitals in a single Slater determinant roughly matched to the low-momentum part of the high-resolution distribution.
In doing so, we sacrifice full consistency for simplicity and wider applicability, without the need for computationally expensive machinery.

High-resolution nucleon momentum distributions are characterized by a ``mean-field'' distribution up to roughly the Fermi momentum joined to a high-momentum tail, with the latter dominantly attributed to SRCs (see Fig.~\ref{fig:proton_momentum_distributions_log}). 
At low RG resolution the wave function becomes increasingly uncorrelated (soft), such that the momentum distribution from matrix elements of the same one-body operator used at high resolution would only exhibit the mean-field part. 
To recover the high-resolution distribution, one includes the
induced two-body operator from RG evolution.
Reference~\cite{Tropiano:2021qgf} demonstrated that a local density approximation (LDA) for the ground-state wave functions was sufficient
to accurately reproduce the high-momentum tail of nucleon momentum distributions, but was not quantitative at lower momenta.
Here we improve upon this approach by using a Slater determinant with phenomenologically fit single-particle (s.p.) wave functions.

Our detailed comparison to high-resolution results with the same short-distance physics uses variational Monte Carlo (VMC) calculations. 
VMC is an \textit{ab initio} nuclear many-body method that solves the Schr{\"o}dinger equation using local interactions in coordinate space~\cite{Carlson:2014vla}.
VMC performs a variational minimization of the stochastically determined energy of a trial-state wave function, which is given by a correlation operator applied to a mean-field state.
In contrast to many other nuclear many-body methods, it can  
accommodate high-resolution interactions and the resulting highly correlated wave functions without difficulties.
Since there is no truncation in the correlation effects, the VMC approach scales exponentially with the number of nucleons, presently limiting its practical applicability to light nuclear systems. 
For larger nuclei, a cluster VMC (CVMC) method has been developed~\cite{Lonardoni:2017egu} in which a full $3A$-dimensional integral is made for the mean-field state, including central SRCs, while a linked  cluster expansion (up to five-body) is made for the spin-isospin correlations.

In Sec.~\ref{sec:methodology} we review the formalism for evaluating SRG-evolved momentum distributions and the methodology for VMC momentum distributions.
In Sec.~\ref{sec:results} SRG distributions evolved from AV18 are compared to VMC and CVMC distributions for AV18 with a three-nucleon potential added, for a range of nuclei.
We find quantitative reproductions of the proton momentum distributions after matching using a single choice of SRG scale $\lambda$, with a clean factorization at low resolution. 
We show the sensitivity to the choice of $\lambda$ and 
also how momentum distributions from other Hamiltonians can be SRG-transformed to compare with AV18 results.
Section~\ref{sec:summary} gives a summary and an outlook with possible extensions of the present work.

\section{Methodology}
\label{sec:methodology}

The single-nucleon momentum distributions at high and low resolution are given by matrix elements in the $A$-nucleon ground state $\ket{\Psi_0^A}$,
\begin{align}
    \label{eq:matrix_element}
    \mel{\Psi_0^A}{\nhathi^{\tau}(\qvec)}{\Psi_0^A}
        &= \mel{\Psi_0^A}{\Uhatlamdag \Uhatlam \nhathi^{\tau}(\qvec) \Uhatlamdag \Uhatlam}{\Psi_0^A} \notag \\
        &\equiv \mel{\Psi_0^A(\lambda)}{\nhatlo^{\tau}(\qvec)}{\Psi_0^A(\lambda)}
     ,
\end{align}
where SRG transformations $\Uhatlam$ are applied to both the wave function and operator~\cite{Tropiano:2021qgf} (after which they are labeled by $\lambda$).
The overall matrix element does not change because SRG transformations are unitary.
Here the initial operator and evolved operators are given in second quantization by
\begin{equation}
    \label{eq:n_infinity}
    \nhathi^{\tau}(\qvec) = 
        \sum_{\sgn}
        \akdag_{\qvec \sgn \tau}
        \ak_{\qvec \sgn \tau}
         ,
\end{equation}
\begin{equation}
    \label{eq:n_lambda}
    \nhatlo^{\tau}(\qvec) = \Uhatlam \nhathi^{\tau}(\qvec) \Uhatlamdag
     ,
\end{equation}
where $\qvec$ is the single-nucleon momentum, $\sgn$ is the spin projection, and $\tau$ is the isospin projection.
The subscript of the operator indicates whether it is SRG-evolved or not ($\lambda = \infty$ is unevolved).
Up to this point, Eq.~\eqref{eq:matrix_element} is exact, regardless of the chosen SRG resolution scale $\lambda$.

We replace the fully evolved ground state $\ket{\Psi_0^A(\lambda)}$ by a single Slater determinant of Woods-Saxon orbitals
\begin{equation}
    \label{eq:low_rg_ground_state}
    \ket{\Psi_0^A(\lambda)} \rightarrow \prod_{\alpha < \rm{F}} \akdag_\alpha \ket{0}
    ,
\end{equation}
where the indices run over occupied s.p.~states $\alpha \equiv (n_\alpha$, $l_\alpha$, $j_\alpha$, $m_{j_\alpha}$, $m_{t_\alpha})$ with spin $s_\alpha = 1/2$ and isospin $t_\alpha=1/2$, and $\rm{F}$ refers to the Fermi surface.
The quantum numbers denoted by $\alpha$ refer to the principal quantum number, orbital angular momentum, total angular momentum, total angular momentum projection, and isospin projection, respectively.
In principle, the low-RG resolution wave function should be obtained by solving the Schr{\"o}dinger equation associated with the evolved Hamiltonian, but in general it should reflect a dominantly ``mean-field'' description of nuclei.
After this approximation, the combination of the wave functions with the evolved operator is no longer unitary, meaning the matrix element will depend on $\lambda$.
The rationale for how $\lambda$ is chosen is given in the following section.

The SRG unitary transformation at flow parameter $\lambda$ has the following schematic form in second quantization:
\begin{equation}
    \label{eq:Uschematic}
    \Uhat_\lambda = \Ihat 
        + \sum  \delta U^{(2)}_\lambda \akdag\akdag aa
        + \sum \delta U^{(3)}_\lambda \akdag\akdag\akdag aaa
        + \cdots
    ,
\end{equation}
where we have suppressed the s.p.~indices and combinatoric factors.
In practice, $\delta U^{(2)}_\lambda$ is obtained in the relative momentum partial-wave basis by solving an SRG flow equation for the transformation directly, given an NN interaction.
We apply SRG transformations to the initial momentum distribution operator~\eqref{eq:n_infinity} and use Wick's theorem in operator form to truncate at the two-body (vacuum) level omitting three-body and higher-body operators of Eq.~\eqref{eq:Uschematic}.
There is an exact cancellation of evolved operators in computing the overall normalization of the nucleon momentum distribution, meaning that proton number $Z$ and neutron number $N$ are preserved~\cite{Tropiano:2021qgf}.

It has been shown that the major features of SRC physics are well described within a two-body approximation~\cite{Tropiano:2021qgf}.
The validity of such an approximation dates back to the seminal work of Brueckner and collaborators~\cite{Brueckner:1955zzd}.
This approximation is also supported by the GCF, which has made a truncation at the two-body level in several calculations~\cite{Weiss:2015mba,Weiss:2016obx,Weiss:2018tbu,Weiss:2021zyb}.
Furthermore, both CVMC and correlated-basis function theory have shown that the two-body cluster contribution is by far the largest in the cluster expansion~\cite{Pandharipande:1979bv,Morales:2002qi,Lonardoni:2017egu}.
We plan to quantify the three-body operator contributions in a future study.
See~\cite{Neff:2015xda,Weiss:2023laq} for further details on three-body contributions from an SRG and GCF standpoint, respectively.

The two-body evolved operator is evaluated with respect to antisymmetrized two-nucleon plane-wave kets \, $\ket{\kvec_1 \sg1 \ta1 \, \kvec_2\sg2\ta2}$, where $\sgn$ and $\tau$ refer to nucleon spin and isospin projections.
Suppressing the momentum, spin, and isospin dependence, the evolved operator has the schematic form
\begin{align}
    \label{eq:evolved_operator}
    \nhatlo^{\tau} &\approx
        \nhathi^{\tau}
        + \sum (\delta U^{(2)}_\lambda \akdag\akdag aa
            + \delta U^{\dagger (2)}_\lambda \akdag\akdag aa) \notag \\
        &\qquad \null + \sum \delta U^{(2)}_\lambda \delta U^{\dagger\,(2)}_\lambda
            \akdag\akdag aa
     .
\end{align}
Equation~\eqref{eq:evolved_operator} is approximate because the three-body and higher operators are truncated.
To evaluate the matrix element,
we transform the creation and annihilation operators from the plane-wave basis to the s.p.~basis of Woods-Saxon orbitals using
\begin{equation}
\label{eq:transformed_annihilation_operator}
    \akfull{\kvec}{\sgn}{\tau} = \sum_\alpha \psi_\alpha(\kvec;\sgn,\tau) \, \ak_\alpha
    ,
\end{equation}
where $\psi_\alpha(\kvec;\sgn,\tau)$ is a s.p.~wave function with respect to orbital $\alpha$, and contractions are given by
\begin{equation}
    \label{eq:contraction}
    \wcA{\alpha}{\beta} = \mel{\Psi_0^A(\lambda)}{\akdag_\alpha \ak_\beta}{\Psi_0^A(\lambda)}
        = \delta_{\alpha \beta}
     ,
\end{equation}
for $\alpha,\beta < \rm{F}$ and zero otherwise.
Following this procedure, the single-nucleon momentum distribution takes the form

\begin{align}
   \label{eq:momentum_distribution}
    n^\tau_\lambda(\qvec) &=
        \sum_{\sgn} \sum_{\alpha < \rm{F}} \abs{\psi_\alpha(\qvec;\sgn,\tau)}^2
  + \frac{1}{2}
        \sum_{\sg{1} \sg{2} \sgn \sgn'}
        \sum_{\ta{1} \ta{2} \tau'}     \sum_{\alpha \beta < \rm{F}}   \notag \\
      &  
       \quad \int d\Kvec \int d\kvec 
        \Big[
            \melr{\kvec \sg{1} \ta{1} \sg{2} \ta{2}}{\delUlam}{\qvec - \Kvec/2 \sgn \tau \sgn' \tau'}
        \notag \\
        & \quad\qquad\times\psi^\dagger_\alpha(\Kvec/2+\kvec;\sg{1},\ta{1})
            \psi^\dagger_\beta(\Kvec/2-\kvec;\sg{2},\ta{2}) \notag \\
            &\quad\qquad \times \bigl(
                \psi_\beta(\Kvec-\qvec;\sgn',\tau')
                \psi_\alpha(\qvec;\sgn,\tau)
           \notag \\ & \qquad\qquad\qquad
                - \psi_\alpha(\Kvec-\qvec;\sgn',\tau')
                \psi_\beta(\qvec;\sgn,\tau)
            \bigr) \notag \\
            &\qquad+ \melr{\qvec - \Kvec/2 \sgn \tau \sgn' \tau'}{\delUdaglam}{\kvec \sg{1} \ta{1} \sg{2} \ta{2}}
            \notag \\ & \quad\qquad \times\psi_\alpha(\Kvec/2+\kvec;\sg{1},\ta{1})
            \psi_\beta(\Kvec/2-\kvec;\sg{2},\ta{2}) \notag \\
            &\quad\qquad \times \bigl(
                \psi^\dagger_\beta(\Kvec-\qvec;\sgn',\tau')
                \psi^\dagger_\alpha(\qvec;\sgn,\tau)
             \notag \\ & \qquad\qquad\qquad   - \psi^\dagger_\alpha(\Kvec-\qvec;\sgn',\tau')
                \psi^\dagger_\beta(\qvec;\sgn,\tau)
            \bigr)
        \Big] \notag \\
        &\quad+ \frac{1}{4}
        \sum_{\sg{1} \sg{2} \sg{3} \sg{4} \sgn \sgn'}
        \sum_{\ta{1} \ta{2} \ta{3} \ta{4} \tau'}
        \sum_{\alpha \beta < \rm{F}}
        \int d\Kvec \int d\kvec \int d\kpvec
        \notag \\ & \qquad\qquad\times
            \melr{\kvec \sg{1} \ta{1} \sg{2} \ta{2}}{\delUlam}{\qvec-\Kvec/2 \sgn \tau \sgn' \tau'} \notag \\
            &\qquad\qquad\times \melr{\qvec-\Kvec/2 \sgn \tau \sgn' \tau'}{\delUdaglam}{\kpvec \sg{3} \ta{3} \sg{4} \ta{4}}
            \notag \\ & \qquad\qquad\times    \psi^\dagger_\alpha(\Kvec/2+\kvec;\sg{1},\ta{1})
                \psi^\dagger_\beta(\Kvec/2-\kvec;\sg{2},\ta{2}) \notag \\
                &\qquad\qquad\times \bigl(
                    \psi_\beta(\Kvec/2-\kpvec;\sg{4},\ta{4})
          \psi_\alpha(\Kvec/2+\kpvec;\sg{3},\ta{3})
               \notag \\ & \quad\qquad
                    - \psi_\alpha(\Kvec/2-\kpvec;\sg{4},\ta{4})
                    \psi_\beta(\Kvec/2+\kpvec;\sg{3},\ta{3})
                \bigr)
     ,
\end{align}
where $\qvec$ is the single-nucleon momentum, $\kvec$ and $\kpvec$ are relative momenta, and $\Kvec$ is the total momentum.
For further details on the derivation of Eq.~\eqref{eq:momentum_distribution}, we refer the reader to the supplemental material.

We benchmark the SRG single-nucleon momentum distributions against those obtained with VMC calculations.
The VMC method \cite{Carlson:2014vla} approximates the solution of the nuclear quantum many-body problem with a variational ansatz of the form
\begin{equation}
    |\Psi_V\rangle = \Big(1 + \sum_{i<j<k}F_{ijk}\Big)\Big(\mathcal{S}\prod_{i<j}F_{ij}\Big) |\Psi_J\rangle\, .
\end{equation}
Here, $F_{ij}$ and $F_{ijk}$ represent two- and three-body correlation operators, respectively, which include both central and spin-isospin-dependent terms; the symbol $\mathcal{S}$ denotes a symmetrized product over nucleon pairs necessary for ensuring permutation invariance in the wave function for those components of $F_{ij}$ which do not commute.
The $\alpha$-cluster  structure of light nuclei is explicitly accounted for by the antisymmetric Jastrow wave function $|\Psi_J\rangle$ that is constructed from a sum over independent-particle terms, each having four nucleons in an $\alpha$-like core and the remaining $(A-4)$ nucleons in p-shell orbitals~\cite{Pieper:2002ne}.
The optimal set of variational parameters, defining $F_{ij}$, $F_{ijk}$, and $|\Psi_J\rangle$, is determined by minimizing the expectation value of the energy:
\begin{equation}
    E_V \equiv \frac{\langle \Psi_V | H | \Psi_V \rangle }{\langle \Psi_V | \Psi_V \rangle } \geq E_0,
\end{equation}
where $E_0$ represents the true ground-state energy of the system, subject to the constraint of obtaining approximately correct charge radii.
Evaluating the above expectation value involves a multi-dimensional integration over the $3A$ spatial coordinates of the nucleons, performed stochastically using the Metropolis-Hastings algorithm \cite{Metropolis:1953,Hastings:1970}.
Conversely, the sum over the $2^A \times \binom{A}{Z}$ spin-isospin coordinates is carried out explicitly, resulting in an exponential cost with the number of nucleons which presently limits the applicability of the VMC to light (up to $^{12}$C) nuclear systems.
For larger systems ($^{16}$O, $^{40}$Ca), the exponential cost can be ameliorated by performing a linked cluster expansion in the spin-isospin-dependent correlations~\cite{Lonardoni:2017egu}.
The $|\Psi_J\rangle$ is now a product of shell-model-like s.p.~determinants and a full $3A$-dimensional integral is evaluated for these and the central parts of $F_{ij}$, which include a substantial part of the SRCs.

We note that as a critical advantage with respect to quantum many-body methods relying on a s.p.~basis expansion, the VMC has no difficulties in dealing with high-resolution nuclear potentials, which generate high-momen\-tum components in the ground-state wave function. 
As discussed in detail in Refs.~\cite{Wiringa:2013ala,Carlson:2014vla}, the VMC momentum distributions are evaluated by
\begin{align}
    n(\mathbf{q}) &= \int d\mathbf{r}_1^\prime d\mathbf{r}_1 d\mathbf{r}_2 \dots d\mathbf{r}_A \Psi^\dagger(\mathbf{r}_1^\prime, \mathbf{r}_2 \dots \mathbf{r}_A) e^{-i \mathbf{q} \cdot (\mathbf{r}_1 - \mathbf{r}_1^\prime)}\nonumber\\
    & \qquad\qquad\times \Psi(\mathbf{r}_1, \mathbf{r}_2 \dots \mathbf{r}_A).
\end{align}
The above Fourier transform is computed by sampling configurations from $|\Psi^\dagger(\mathbf{r}_1, \mathbf{r}_2 \dots \mathbf{r}_A)|^2$.
We average over all particles $i$ in each configuration, and for each particle, a grid of Gauss-Legendre points along a random direction is used to compute the Fourier transform.
To reduce the statistical errors originating from the rapidly oscillating nature of the integrand, instead of just moving the position $\mathbf{r}_i^\prime$ in the left-hand wave function away from a fixed position $\mathbf{r}_i$ in the right-hand wave function, both positions are moved symmetrically away from $\mathbf{r}_i$.
The VMC wave functions reproduce experimental charge radii of $^4$He and $^{12}$C within $\sim$1\%, while the kinetic energy matches the more precise Green's function Monte Carlo calculations within $\sim$2-8\%.
Because nearly half the kinetic energy comes from the high-momentum tails~\cite{Wiringa:single_distributions}, we believe these VMC momentum distributions to be slightly low but fairly accurate.
The CVMC radii for $^{16}$O and $^{40}$Ca are $\sim$2-4\% larger than experiment.

\section{Results}
\label{sec:results}

\begin{figure}[tbh]
    \centering
    \includegraphics[width=0.8\columnwidth]{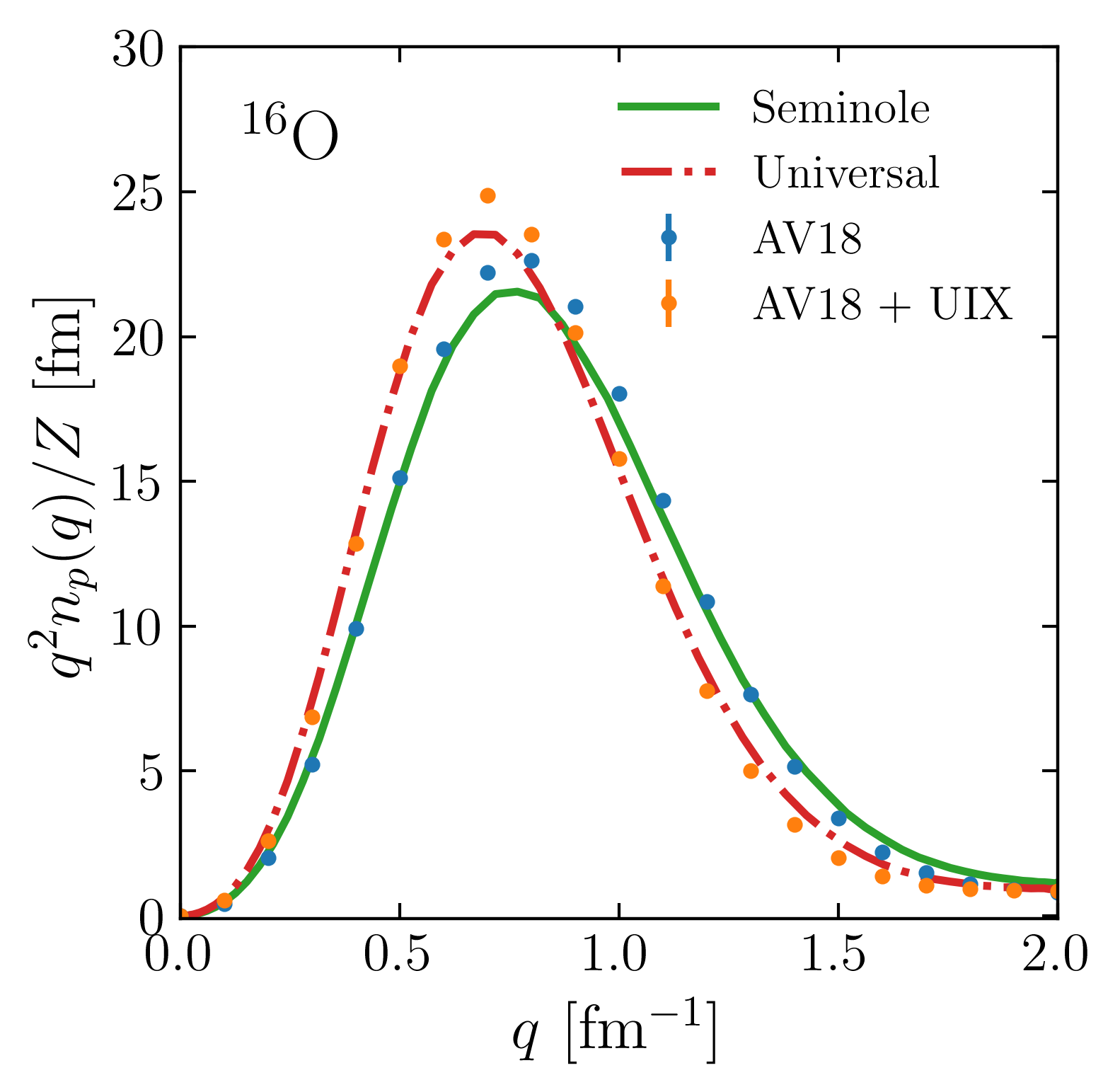}
    \caption{%
    Proton momentum distributions of \oxygen\, with different Woods-Saxon parametrizations compared to CVMC results with and without a three-nucleon interaction.
    The SRG distributions are calculated using the AV18 interaction to evolve the operator to $\lambda=1.5\,\fmi$, comparing the Seminole parametrization (solid green) and Universal parametrization (dashdotted red).
    The CVMC results are calculated either with AV18 only (blue points), or AV18 and the Urbana IX (UIX) three-nucleon potential (orange points).
    }
    \label{fig:O16_ws_comparison}
\end{figure}

In Fig.~\ref{fig:O16_ws_comparison} we show single-proton high-resolution momentum distributions up to $2\,\fmi$ for \oxygen\ from CVMC calculations using both the AV18 two-nucleon interaction only and with AV18 and the Urbana IX (UIX) three-nucleon potential~\cite{Pudliner:1995wk}.
The differences between the distributions are typical of other nuclei and also of differences between using UIX and the Urbana X (UX) potential~\cite{Wiringa:2013ala}.
To match to SRG distributions using Eq.~\eqref{eq:momentum_distribution} we need to choose an appropriate s.p.~basis for each interaction.
We use adjusted Woods-Saxon orbitals 
as a phenomenological way to build in the relevant nuclear saturation physics without having to explicitly evolve three- and higher-body forces.

Figure \ref{fig:O16_ws_comparison} shows SRG results with two different Woods-Saxon parametrizations dubbed ``Universal''~\cite{Dudek:1982zz} and ``Seminole''~\cite{Schwierz:2007ve,Volya:woods_saxon}, with both using AV18 to SRG-evolve the operator to $\lambda=1.5\,\fmi$.
The Universal param\-etrization describes heavy nuclei such as $^{208}$Pb, whereas the Seminole parametrization is intended for shell model calculations of \oxygen\ and heavier nuclei.
The Universal distribution tends to be close to the CVMC distribution with AV18 and UIX, while the Seminole favors the AV18-only distribution.
In all subsequent figures, we adjust the strength and radius of the central
potential to best describe VMC calculations using the same values as the Universal
parametrization for the other Woods-Saxon parameters; however, no fine-tuning of the parameters is necessary.

\begin{figure*}[tbh]
    \centering
    \includegraphics[clip,width=0.98\textwidth]{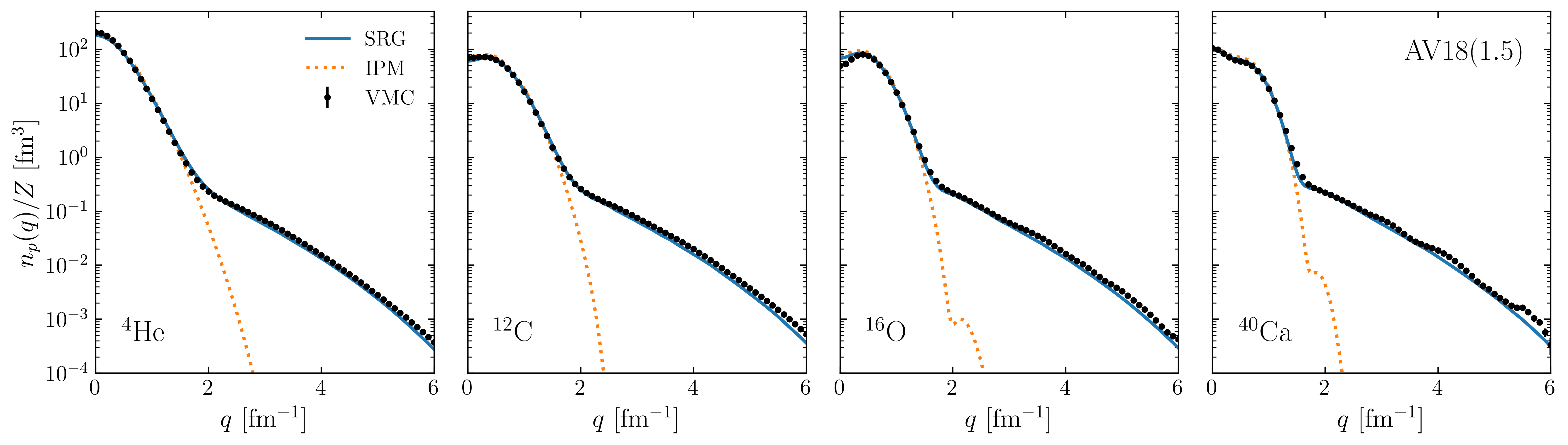}
    \caption{%
    Proton momentum distributions for \helium, \carbon, \oxygen, and \calcium.
    The solid blue lines show the SRG distributions in which the operator is evolved under the AV18 interaction at $\lambda=1.5\,\fmi$.
    The dashed orange lines show the IPM distributions (no operator evolution).
    The black points show VMC distributions calculated with AV18 and UX for \helium\ and \carbon, and CVMC distributions calculated with AV18 and UIX for \oxygen\ and \calcium.
    Each distribution is divided by the proton number $Z$.
    }
    \label{fig:proton_momentum_distributions_log}
    \bigskip
    \centering
    \includegraphics[clip,width=0.98\textwidth]{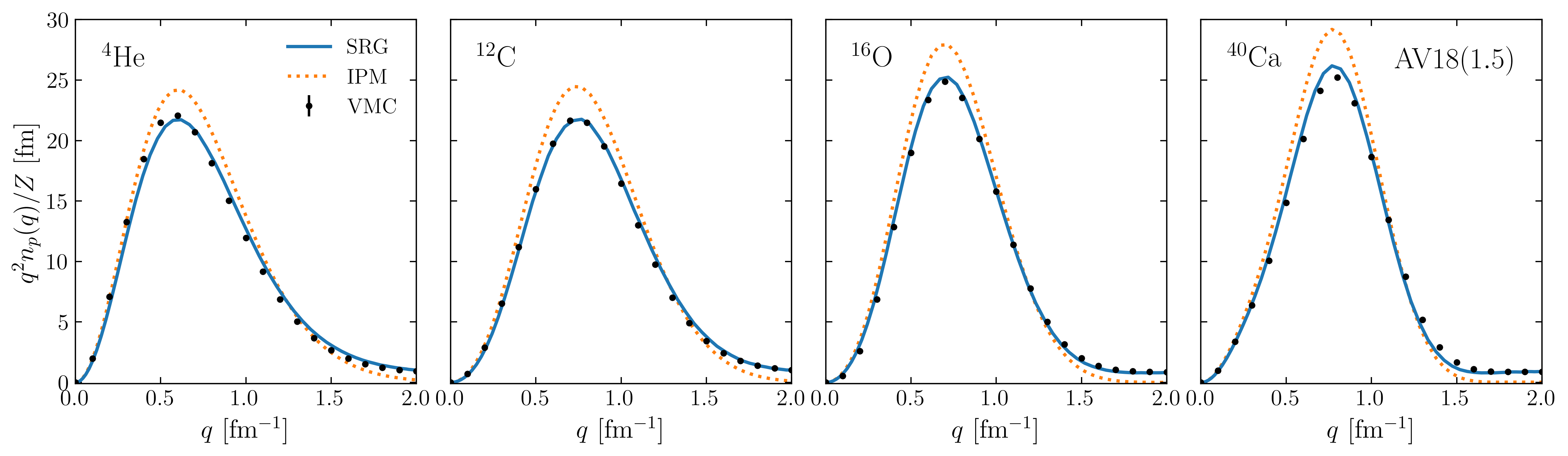}
    \caption{%
    Same as Fig.~\ref{fig:proton_momentum_distributions_log} but on a linear y-scale with a factor of $q^2$ included.
    }
    \label{fig:proton_momentum_distributions_linear}
\end{figure*}

Note that if one tried to use consistently calculated Hartree-Fock orbitals with soft NN-only Hamiltonians, 
the results would be poor because saturation would be distorted.
The semi-phenomenological approach is in the spirit of hybrid calculations that successfully mix accurate structure with effective field theory current operators 
as well as with the traditional phenomenological analysis of $(e,e'p)$ experiments~\cite{Lapikas:1993aa}. 
Our justification here is empirical but is open to more controlled validation through further benchmarking with {\it ab initio} many-body calculations.

Spurious center-of-mass (CoM) effects can be sizable in calculations involving a Woods-Saxon s.p.~basis for light nuclei.
Subtracting the spurious CoM effect from calculated wave functions is a nontrivial issue.
However the VMC performs a Monte Carlo integration in which the CoM component is exactly subtracted from the wave function~\cite{Massella:2018xdj}.
We have compared VMC calculations with and without the CoM subtraction for \helium\, and \carbon\, finding a sizable effect in the former.
Without the subtraction, the single-nucleon momentum distribution for \helium\, is shifted to higher momentum, meaning that the probability of finding a nucleon with low (high) momentum decreases (increases).
This is due to the spurious CoM motion giving an overall enhancement to the kinetic energy.
The low RG resolution calculations do not make any CoM subtraction because the Woods-Saxon potential is adjusted to match the CoM-subtracted VMC distributions.

Figure~\ref{fig:proton_momentum_distributions_log} shows SRG proton momentum distributions of $^4$He, $^{12}$C, $^{16}$O, and $^{40}$Ca from Eq.~\eqref{eq:momentum_distribution} using AV18 with $\lambda = 1.5\,\fmi$ compared to VMC and CVMC results.
The figure label AV18(1.5) refers to the AV18 potential with $\lambda = 1.5\,\fmi$ for all SRG calculations shown in this section.
VMC with AV18 and UX is used to calculate \helium\ and \carbon, and CVMC with AV18 and UIX is used for \oxygen\ and \calcium.
See the supplemental material for comparisons of SRG to VMC (or CVMC) with AV18 only.
The orange dotted lines correspond to a single Slater determinant of Woods-Saxon s.p.~states adjusted to either VMC results with AV18 and UX, or CVMC with AV18 and UIX.
Including operator evolution reduces the independent particle model (IPM) description by negative $\delta U$ and $\delta U^\dagger$ linear contributions as seen in the linear y-scale $q^2 n(q)$ figure~\ref{fig:proton_momentum_distributions_linear}.
The high momentum tail arises from the $\delta U \delta U^\dagger$ two-body term dependent on the NN interaction.
The tail agrees nicely with the VMC and CVMC calculations regardless of nuclei because each calculation uses the same two-nucleon interaction AV18, which is the dominant contribution at high momentum.

\begin{figure}[!htb]
    \centering
    \includegraphics[width=0.8\columnwidth]{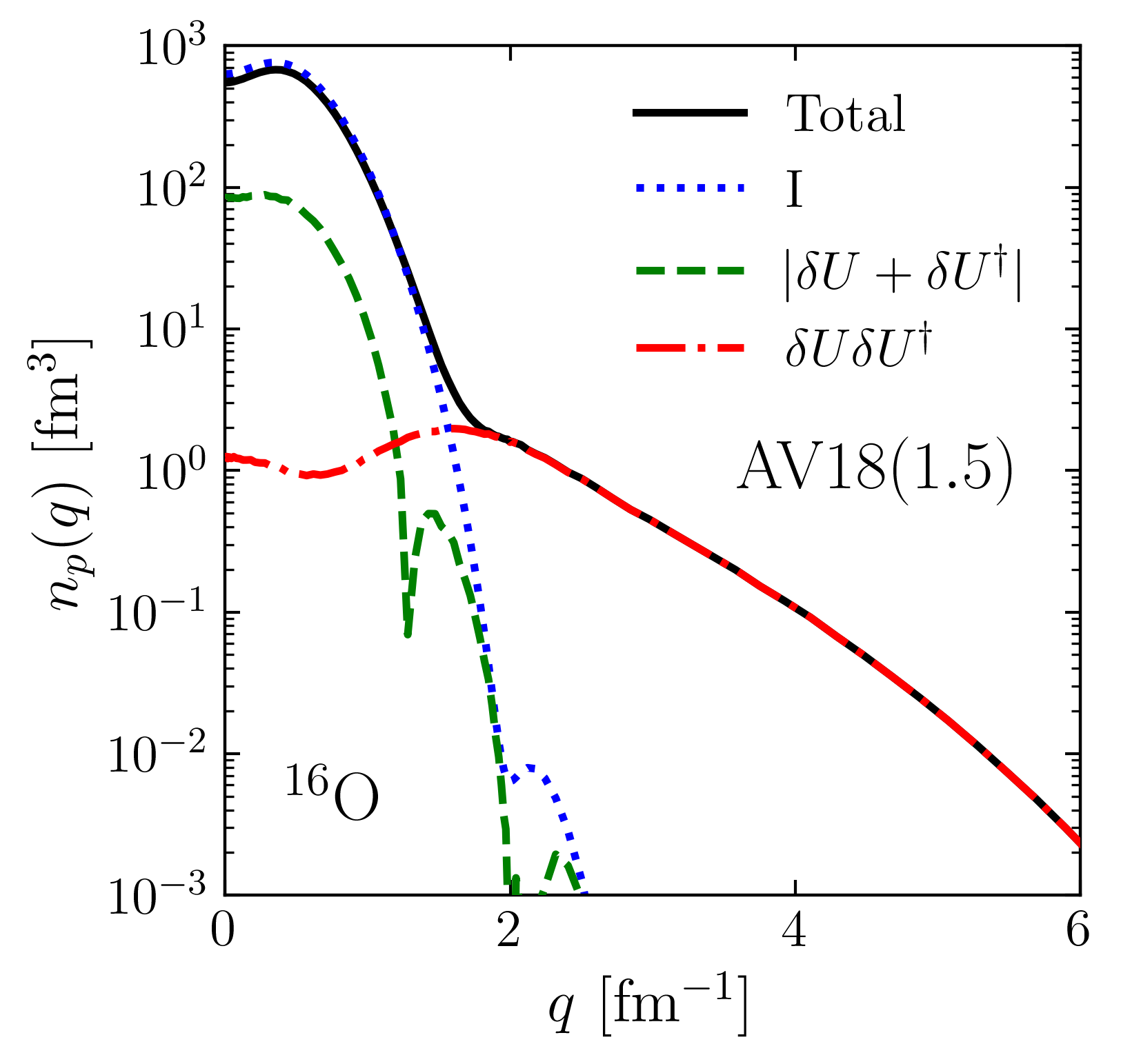}
    \caption{%
    Contributions to the proton momentum distribution in \oxygen, corresponding to the first (mean-field), second ($\delta U + \delta U^\dagger$), and third ($\delta U\delta U^\dagger$) terms in Eq.~\eqref{eq:momentum_distribution}, all evolved to SRG $\lambda = 1.5\,\fmi$ with AV18. 
    The sum of the three is the solid line. 
    Note that the dashed $\delta U + \delta U^\dagger$ contribution is negative up to $q \sim 1.4\,\fmi$.
    }
    \label{fig:o16_proton_momentum_distribution_contributions}
\end{figure}

Figure~\ref{fig:o16_proton_momentum_distribution_contributions} shows the contributions to the \oxygen\ SRG proton momentum distribution.
The black solid line shows the total momentum distribution, the blue dotted shows the contribution from the unevolved operator (i.e., IPM), the green dashed line shows the absolute value of the $\delta U$ and $\delta U^\dagger$ terms, and the red dash-dotted line shows the $\delta U \delta U^\dagger$ term.
The $\delta U + \delta U^\dagger$ contribution is negative up to about $1.4\,\fmi$ reducing (``quenching'') the distribution from the IPM.
The IPM and $\delta U + \delta U^\dagger$ contributions are weighted by s.p.~wave functions that carry the $q$ dependence.
These wave functions do not have high momentum components, hence the two contributions drop off at high $q$.
The $q$ dependence of the $\delta U \delta U^\dagger$ contribution is entirely driven by the $\delta U$ and $\delta U^\dagger$ matrix elements and gives the full contribution to the distribution at high $q$.
This contribution corresponds to the tail at high resolution originating with pairs in the Fermi sea being kicked to high momentum by a hard interaction and then dropping back with another interaction~\cite{Brueckner:1955zzd}.

\begin{figure}[!htb]
    \centering
    \includegraphics[width=0.8\columnwidth]{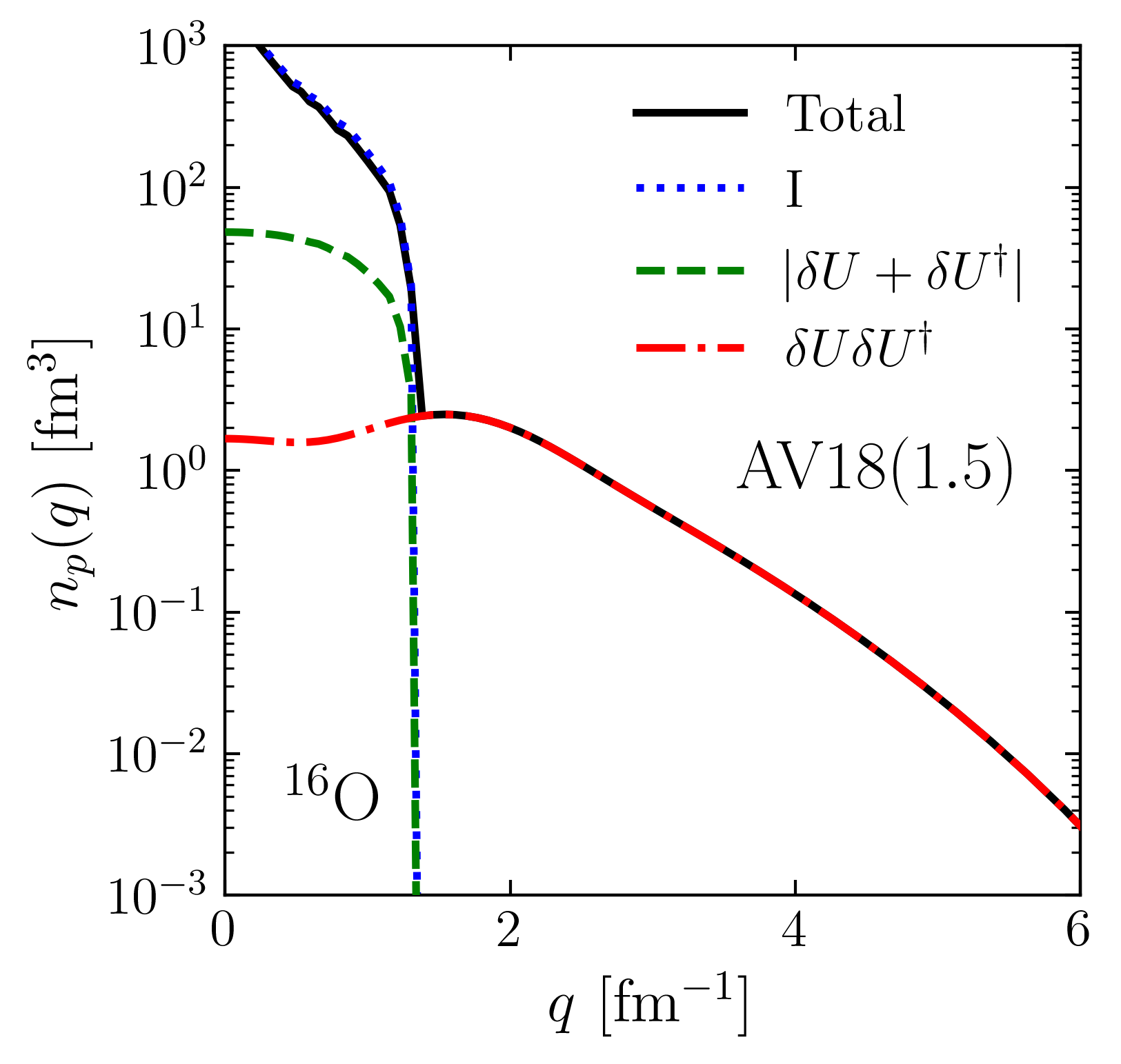}
    \caption{%
    Same as Fig.~\ref{fig:o16_proton_momentum_distribution_contributions} but using a local density approximation to compute proton momentum distributions as described in Ref.~\cite{Tropiano:2021qgf}.
    Here the proton densities are generated from the SLy4 Skyrme functional~\cite{Chabanat:1997un} using the HFBRAD code~\cite{Bennaceur:2005mx}.
    }
    \label{fig:o16_proton_momentum_distribution_contributions_lda}
\end{figure}

The high momentum tail is explained by factorization of SRG transformations when there is a separation of scales~\cite{Anderson:2010aq,Bogner:2012zm}.
Mathematically, $\delUlam(k,q) \approx \Flo_\lambda(k) \Fhi_\lambda(q)$ for $k < \lambda \ll q$, where the labels ``hi'' and ``lo'' in the functions $\Fhi_\lambda(q)$ and $\Flo_\lambda(k)$ refer to the separation of momentum scales above and below $\lambda$.
The low RG resolution wave function only supports momenta up to the Fermi momentum, which is generally less than $\lambda=1.5\,\fmi$ for all nuclei considered in this paper.
Thus at high $q$, the $\delta U \delta U^\dagger$ term factorizes into a universal two-body function $\abs{\Fhi_\lambda(q)}^2$ that depends on the interaction but not the nucleus, and a low momentum nuclear matrix element independent of the interaction and $q$:
\begin{equation}
    \label{eq:delU2_term}
    \lim_{q \gg \lambda} n_\lambda(q) \propto
        \abs{\Fhi_\lambda(q)}^2 \int \mel{A}{\Flo_\lambda(k)\Flo_\lambda(k')}{A}.
\end{equation}
This implies scaling of high-momentum tails because the high-$q$ dependence cancels in ratios of nuclei leaving a quantity only sensitive to low momentum physics (e.g., SRC scaling factors $a_2$).
Note the universal high-momentum tail at $q \gg \lambda$ in the proton momentum distributions in Fig.~\ref{fig:proton_momentum_distributions_log}.

Figure~\ref{fig:o16_proton_momentum_distribution_contributions_lda} shows the contributions to the proton momentum distribution but using a local density approximation (LDA) to model the low RG resolution wave function.
This approach is described in Ref.~\cite{Tropiano:2021qgf} and was also utilized in Ref.~\cite{Tropiano:2022jjj}.
The distribution in Fig.~\ref{fig:o16_proton_momentum_distribution_contributions_lda} uses proton densities generated from the SLy4 Skyrme functional~\cite{Chabanat:1997un} using the HFBRAD code~\cite{Bennaceur:2005mx}.
The mean-field part of the momentum distribution will reflect the differences in modeling $\ket{\Psi_0^A(\lambda)}$, despite the SRG transformations and factorization arguments remaining the same.
In the LDA version, there are sharp cutoffs at the SRG resolution scale $\lambda=1.5\,\fmi$, whereas the Woods-Saxon version smoothly falls off before transitioning to the $\delta U \delta U^\dagger$ term.
The LDA distributions also diverge as $q \rightarrow 0\,\fmi$ due to it being a poor approximation at low momentum.
However, at $q \gg \lambda$ where factorization holds, we retain the same high $q$ tail regardless of LDA or Woods-Saxon.

\begin{figure}[tbh]
    \centering
    \includegraphics[width=0.8\columnwidth]{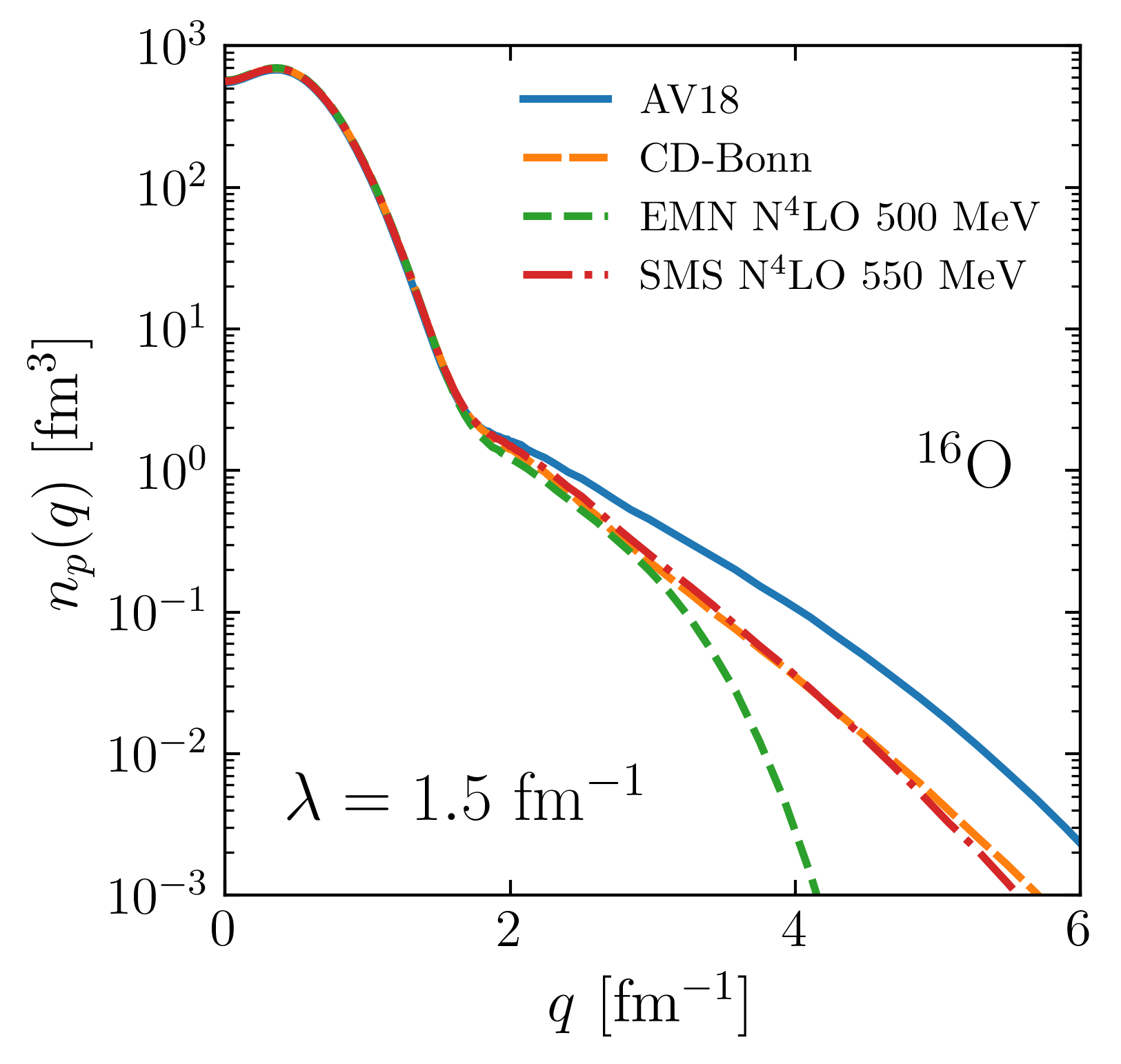}
    \caption{%
    Proton momentum distributions of \oxygen\, using several interactions shown in the legend evolved to $\lambda=1.5\,\fmi$.
    }
    \label{fig:O16_proton_momentum_distribution_vary_kvnn}
\end{figure}

\begin{figure}[tbh]
   \centering     \includegraphics[clip,width=0.8\columnwidth]{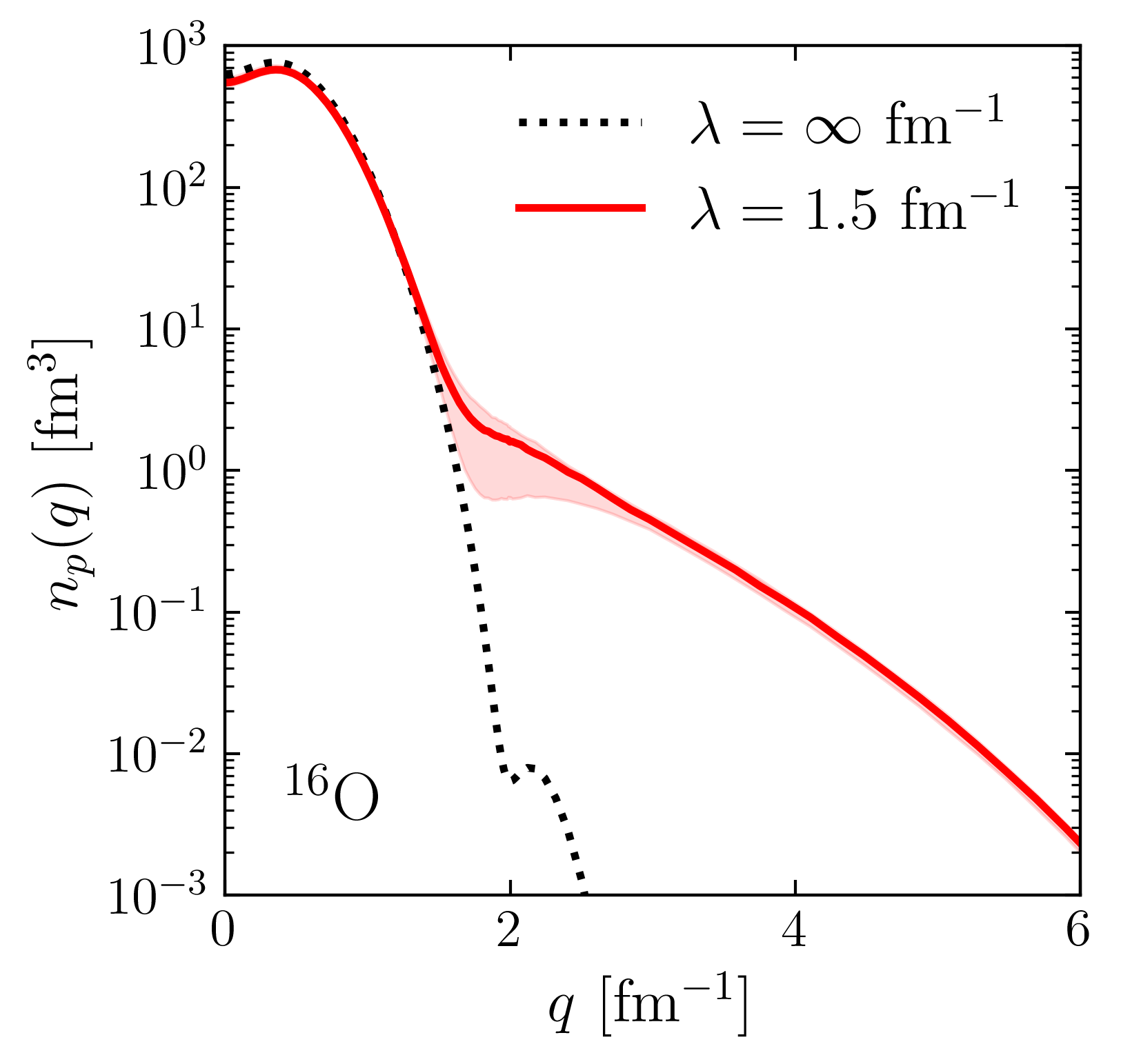}
  \medskip	\includegraphics[clip,width=0.8\columnwidth]{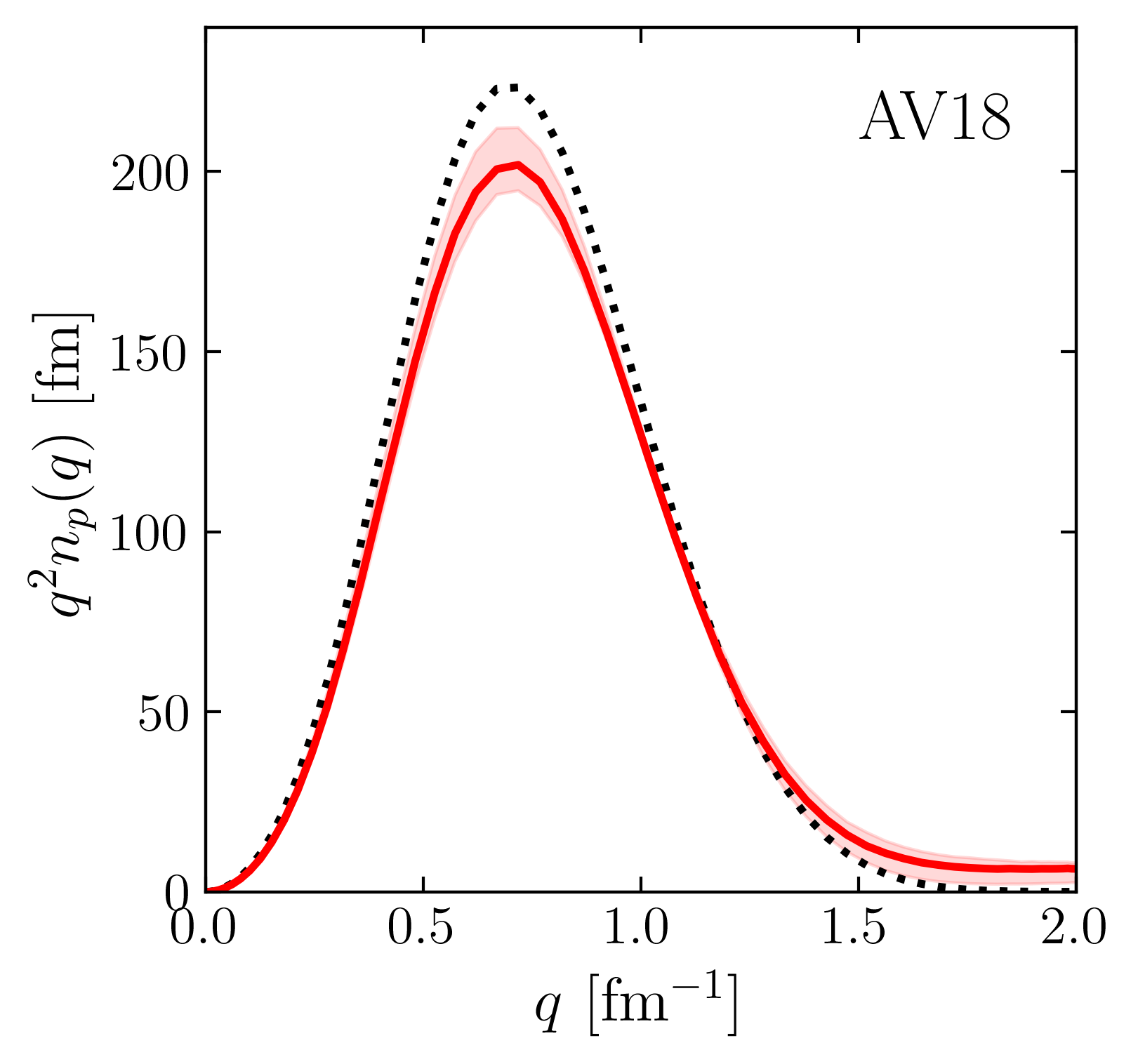}%
    \caption{%
    Proton momentum distribution for \oxygen\, varying the SRG $\lambda$ on a log y-scale (a) and linear y-scale (b) using the AV18 interaction.
    The black dotted line shows the distribution with no SRG evolution (IPM distribution), and the red solid line shows the distribution evolved to $\lambda=1.5\,\fmi$.
    The red band indicates the variation in $\lambda$ from $2$ to $1.35\,\fmi$.
    }
	\label{fig:O16_proton_momentum_distribution_vary_lambda}
\end{figure}
\begin{figure}[!tbh]
    \centering
    \includegraphics[width=0.8\columnwidth]{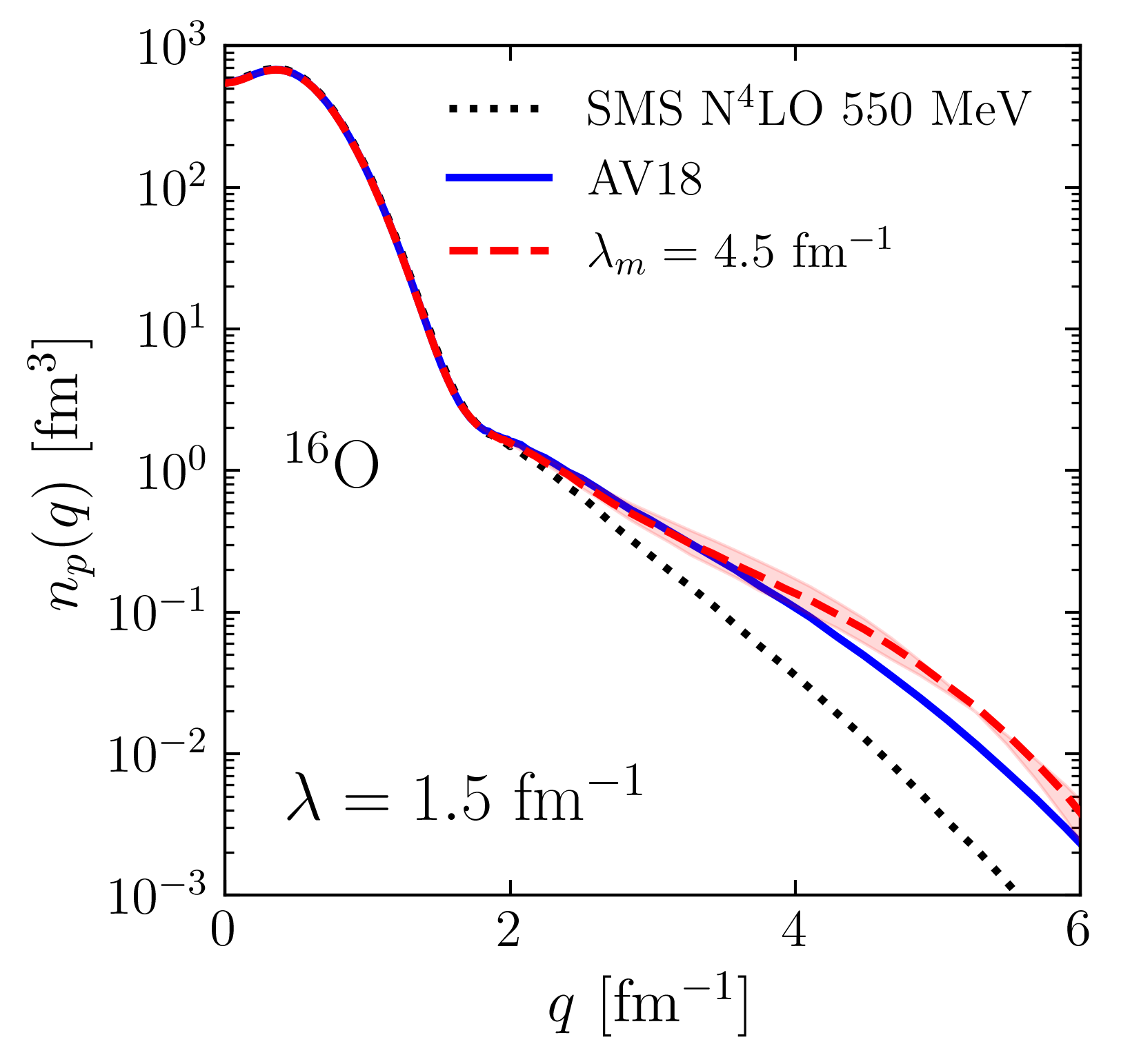}
    \caption{%
    Proton momentum distributions using the matching procedure described in the text to match SMS N$^4$LO 550 MeV to AV18.
    Here we calculate \oxygen\, setting $\lambda=1.5\,\fmi$ in all cases.
    The black dotted line shows the SMS N$^4$LO 550 MeV distribution without matching, and the solid blue line shows the AV18 distribution.
    The red dashed line shows the SMS N$^4$LO 550 MeV distribution with $\lambda_m=4.5\,\fmi$ to match to AV18.
    The red band indicates variation in $\lambda_m$ from $5$ to $4\,\fmi$.
    }
    \label{fig:O16_matching_distribution}
\end{figure}

In Fig.~\ref{fig:O16_proton_momentum_distribution_vary_kvnn} we show SRG proton momentum distributions for \oxygen\ with different NN interactions.
We plot momentum distributions corresponding to two phenomenological interactions, AV18 and CD-Bonn~\cite{Machleidt:2000ge}, and two chiral effective field theory interactions, EMN \NNNNLO\ 500\,MeV \cite{Entem:2017gor} and SMS \NNNNLO\ 550\,MeV \cite{Reinert:2017usi}. 
The $\delta U$ matrix elements in Eq.~\eqref{eq:momentum_distribution} change between each curve because of a different interaction including different regulator schemes.
Variation in the potential has the most visible effect on the high momentum tails because of the dominant $\delta U \delta U^\dagger$ term.
There is also an effect at low momentum where the IPM distribution is quenched by some amount dependent on the interaction.
Figure~\ref{fig:O16_proton_momentum_distribution_vary_kvnn} reflects the scale and scheme dependence associated with the choice of the NN interaction in combination with SRG evolving the same initial one-body operator for the distributions.

In Fig.~\ref{fig:O16_proton_momentum_distribution_vary_lambda} we vary the SRG scale $\lambda$.
The black dotted line shows the result using the unevolved operator, that is, the IPM distribution.
The red line shows the $\lambda=1.5\,\fmi$ distribution.
The light red band indicates the variation in $\lambda$ from $2\,\fmi$ down to $1.35\,\fmi$.
High $\lambda$ distributions approach the IPM description due to the mismatch of a low RG resolution wave function with a high-resolution operator.
As $\lambda$ is lowered, the tail rises from the induced two-body operator, and the low-momentum part of the distribution begins to decrease from the IPM.

Reference~\cite{Tropiano:2022jjj} demonstrates how to use SRG transformations to approximately match different interactions.
Here we match distributions from different interactions to the AV18 distribution, where we use a one-body operator for the AV18 distribution.
Figure \ref{fig:O16_matching_distribution} shows the \oxygen\, proton momentum distributions for SMS N$^4$LO 550 MeV and AV18, where the former has less contributions at high relative momentum.
We apply SRG transformations from AV18 onto the initial momentum distribution operator for the SMS N$^4$LO 550 MeV distribution: $\nhat_{\rm{soft}} = \Uhatlam \nhat_{\rm{hard}} \Uhatlamdag$.
The initial operator for the soft potential $\nhat_{\rm{soft}}$ becomes a two-body operator at some SRG scale indicated by $\lambda_m$.
We then apply the same method as before in evolving the operator and approximating the low RG resolution wave function.
Figure \ref{fig:O16_matching_distribution} shows the SMS N$^4$LO 550 MeV with the initial two-body operator at $\lambda_m=4.5\,\fmi$, where the red band indicates $\lambda_m$ varying from $5$ to $4\,\fmi$.
The induced two-body operator in the initial operator is responsible for approximately matching the SMS N$^4$LO 550 MeV distribution to the AV18 distribution.
We have verified that other matching procedures work in a similar way, such as using unitary transformations that directly relate the eigenstates of both potentials.

\section{Summary and outlook}
\label{sec:summary}

We have demonstrated in this paper that momentum distributions of the high-resolution AV18 potential for nuclei from $A=4$ to $A=40$ can be quantitatively reproduced at low- and high-momenta using an SRG-evolved operator truncated at the two-body level combined with structure described by Slater determinants of adjusted Woods-Saxon orbitals.
The sensitivity to the choice of SRG flow parameter $\lambda$ is relatively small, and the same ``optimal'' choice works for all the nuclei.
The enhanced factorization of the low-resolution operators is reflected in a clean separation of mean-field and SRC contributions to the momentum distributions.
We also show that other interactions, such as state-of-the-art chiral EFT NN forces, can be matched to the reference AV18 distributions to reproduce close to the same distributions.

A key question is whether these successes can be extended to other operators, and ultimately to comparisons with experiment.
At the same time, we need to quantify the error introduced by both the truncation of induced many-body components in the operators and the use of simplified many-body wave functions (and whether we are exploiting cancellations between these errors).
Work is in progress to examine pair momentum distributions as well as exclusive quantities, such as spectral functions for analyzing $(e,e'p)$ experiments at low resolution.
The latter is a doorway to more controlled phenomenological analysis and a modern RG-based formulation of optical potentials~\cite{Hisham:2022jzt} and spectroscopic factors.
We will also evolve operators in single- and double-beta decay calculations to understand quenching from a low RG resolution perspective and connect to GCF approaches~\cite{Weiss:2021rig}.

\section{Acknowledgments}
This work was supported in part by the U.S. Department of Energy (DOE), Office of Science, Office of Nuclear Physics under contract number DE-AC02-06CH11357 (A.J.T., A.L., and R.B.W.) and by the SciDAC-NUCLEI (A.J.T., A.L., R.B.W., R.J.F., and M.A.H.) and SciDAC-NeuCol (A.L.) projects.
A.L. is also supported by DOE Early Career Research Program awards.
R.J.F. and M.A.H. are also supported in part by the National Science Foundation Award No.~PHY-2209442.
S.K.B. is supported in part by the National Science Foundation Award No.~PHY-2310020.
Numerical calculations were performed on the parallel computers of the Laboratory Computing Resource Center, Argonne National Laboratory, the computers of the Argonne Leadership Computing Facility via the INCITE grant ``Ab-initio Nuclear Structure and Nuclear Reactions''.

\bibliographystyle{elsarticle-num} 
\biboptions{sort&compress}
\bibliography{tropiano_bib}

\begin{thebibliography}{10}
\expandafter\ifx\csname url\endcsname\relax
  \def\url#1{\texttt{#1}}\fi
\expandafter\ifx\csname urlprefix\endcsname\relax\def\urlprefix{URL }\fi
\expandafter\ifx\csname href\endcsname\relax
  \def\href#1#2{#2} \def\path#1{#1}\fi

\bibitem{Tropiano:2021qgf}
A.~J. Tropiano, S.~K. Bogner, R.~J. Furnstahl, {Short-range correlation physics
  at low renormalization group resolution}, Phys. Rev. C 104~(3) (2021) 034311.
\newblock \href {http://arxiv.org/abs/2105.13936} {\path{arXiv:2105.13936}},
  \href {https://doi.org/10.1103/PhysRevC.104.034311}
  {\path{doi:10.1103/PhysRevC.104.034311}}.

\bibitem{Bogner:2006pc}
S.~K. Bogner, R.~J. Furnstahl, R.~J. Perry, {Similarity Renormalization Group
  for Nucleon-Nucleon Interactions}, Phys. Rev. C 75 (2007) 061001.
\newblock \href {http://arxiv.org/abs/nucl-th/0611045}
  {\path{arXiv:nucl-th/0611045}}, \href
  {https://doi.org/10.1103/PhysRevC.75.061001}
  {\path{doi:10.1103/PhysRevC.75.061001}}.

\bibitem{Bogner:2009bt}
S.~K. Bogner, R.~J. Furnstahl, A.~Schwenk, {From low-momentum interactions to
  nuclear structure}, Prog. Part. Nucl. Phys. 65 (2010) 94--147.
\newblock \href {http://arxiv.org/abs/0912.3688} {\path{arXiv:0912.3688}},
  \href {https://doi.org/10.1016/j.ppnp.2010.03.001}
  {\path{doi:10.1016/j.ppnp.2010.03.001}}.

\bibitem{Furnstahl:2013oba}
R.~J. Furnstahl, K.~Hebeler, {New applications of renormalization group methods
  in nuclear physics}, Rept. Prog. Phys. 76 (2013) 126301.
\newblock \href {http://arxiv.org/abs/1305.3800} {\path{arXiv:1305.3800}},
  \href {https://doi.org/10.1088/0034-4885/76/12/126301}
  {\path{doi:10.1088/0034-4885/76/12/126301}}.

\bibitem{Hergert:2016iju}
H.~Hergert, S.~K. Bogner, J.~G. Lietz, T.~D. Morris, S.~Novario, N.~M.
  Parzuchowski, F.~Yuan, {In-Medium Similarity Renormalization Group Approach
  to the Nuclear Many-Body Problem}, Lect. Notes Phys. 936 (2017) 477--570.
\newblock \href {http://arxiv.org/abs/1612.08315} {\path{arXiv:1612.08315}},
  \href {https://doi.org/10.1007/978-3-319-53336-0_10}
  {\path{doi:10.1007/978-3-319-53336-0_10}}.

\bibitem{Wiringa:1994wb}
R.~B. Wiringa, V.~Stoks, R.~Schiavilla, {An Accurate nucleon-nucleon potential
  with charge independence breaking}, Phys. Rev. C 51 (1995) 38--51.
\newblock \href {http://arxiv.org/abs/nucl-th/9408016}
  {\path{arXiv:nucl-th/9408016}}, \href
  {https://doi.org/10.1103/PhysRevC.51.38} {\path{doi:10.1103/PhysRevC.51.38}}.

\bibitem{Carlson:1983kq}
J.~Carlson, V.~R. Pandharipande, R.~B. Wiringa, {Three-nucleon interaction in
  3-body, 4-body, and infinite-body systems}, Nucl. Phys. A 401 (1983) 59--85.
\newblock \href {https://doi.org/10.1016/0375-9474(83)90336-6}
  {\path{doi:10.1016/0375-9474(83)90336-6}}.

\bibitem{Pieper:2001ap}
S.~C. Pieper, V.~R. Pandharipande, R.~B. Wiringa, J.~Carlson, {Realistic models
  of pion exchange three nucleon interactions}, Phys. Rev. C 64 (2001) 014001.
\newblock \href {http://arxiv.org/abs/nucl-th/0102004}
  {\path{arXiv:nucl-th/0102004}}, \href
  {https://doi.org/10.1103/PhysRevC.64.014001}
  {\path{doi:10.1103/PhysRevC.64.014001}}.

\bibitem{Carlson:2014vla}
J.~Carlson, S.~Gandolfi, F.~Pederiva, S.~C. Pieper, R.~Schiavilla, K.~E.
  Schmidt, R.~B. Wiringa, {Quantum Monte Carlo methods for nuclear physics},
  Rev. Mod. Phys. 87 (2015) 1067.
\newblock \href {http://arxiv.org/abs/1412.3081} {\path{arXiv:1412.3081}},
  \href {https://doi.org/10.1103/RevModPhys.87.1067}
  {\path{doi:10.1103/RevModPhys.87.1067}}.

\bibitem{Lonardoni:2017egu}
D.~Lonardoni, A.~Lovato, S.~C. Pieper, R.~B. Wiringa, {Variational calculation
  of the ground state of closed-shell nuclei up to $A=40$}, Phys. Rev. C 96~(2)
  (2017) 024326.
\newblock \href {http://arxiv.org/abs/1705.04337} {\path{arXiv:1705.04337}},
  \href {https://doi.org/10.1103/PhysRevC.96.024326}
  {\path{doi:10.1103/PhysRevC.96.024326}}.

\bibitem{Weiss:2015mba}
R.~Weiss, B.~Bazak, N.~Barnea, {Generalized nuclear contacts and momentum
  distributions}, Phys. Rev. C 92~(5) (2015) 054311.
\newblock \href {http://arxiv.org/abs/1503.07047} {\path{arXiv:1503.07047}},
  \href {https://doi.org/10.1103/PhysRevC.92.054311}
  {\path{doi:10.1103/PhysRevC.92.054311}}.

\bibitem{Weiss:2016obx}
R.~Weiss, R.~Cruz-Torres, N.~Barnea, E.~Piasetzky, O.~Hen, {The nuclear
  contacts and short range correlations in nuclei}, Phys. Lett. B 780 (2018)
  211--215.
\newblock \href {http://arxiv.org/abs/1612.00923} {\path{arXiv:1612.00923}},
  \href {https://doi.org/10.1016/j.physletb.2018.01.061}
  {\path{doi:10.1016/j.physletb.2018.01.061}}.

\bibitem{Weiss:2018tbu}
R.~Weiss, I.~Korover, E.~Piasetzky, O.~Hen, N.~Barnea, {Energy and momentum
  dependence of nuclear short-range correlations - Spectral function, exclusive
  scattering experiments and the contact formalism}, Phys. Lett. B 791 (2019)
  242--248.
\newblock \href {http://arxiv.org/abs/1806.10217} {\path{arXiv:1806.10217}},
  \href {https://doi.org/10.1016/j.physletb.2019.02.019}
  {\path{doi:10.1016/j.physletb.2019.02.019}}.

\bibitem{Weiss:2021zyb}
R.~Weiss, A.~W. Denniston, J.~R. Pybus, O.~Hen, E.~Piasetzky, A.~Schmidt, L.~B.
  Weinstein, N.~Barnea, {Extracting the number of short-range correlated
  nucleon pairs from inclusive electron scattering data}, Phys. Rev. C 103~(3)
  (2021) L031301.
\newblock \href {http://arxiv.org/abs/2005.01621} {\path{arXiv:2005.01621}},
  \href {https://doi.org/10.1103/PhysRevC.103.L031301}
  {\path{doi:10.1103/PhysRevC.103.L031301}}.

\bibitem{Hen:2016kwk}
O.~Hen, G.~Miller, E.~Piasetzky, L.~Weinstein, {Nucleon-Nucleon Correlations,
  Short-lived Excitations, and the Quarks Within}, Rev. Mod. Phys. 89~(4)
  (2017) 045002.
\newblock \href {http://arxiv.org/abs/1611.09748} {\path{arXiv:1611.09748}},
  \href {https://doi.org/10.1103/RevModPhys.89.045002}
  {\path{doi:10.1103/RevModPhys.89.045002}}.

\bibitem{Korover:2014dma}
I.~Korover, et~al., {Probing the Repulsive Core of the Nucleon-Nucleon
  Interaction via the $^4He(e,e'pN)$ Triple-Coincidence Reaction}, Phys. Rev.
  Lett. 113~(2) (2014) 022501.
\newblock \href {http://arxiv.org/abs/1401.6138} {\path{arXiv:1401.6138}},
  \href {https://doi.org/10.1103/PhysRevLett.113.022501}
  {\path{doi:10.1103/PhysRevLett.113.022501}}.

\bibitem{Hen:2014nza}
O.~Hen, et~al., {Momentum sharing in imbalanced Fermi systems}, Science 346
  (2014) 614--617.
\newblock \href {http://arxiv.org/abs/1412.0138} {\path{arXiv:1412.0138}},
  \href {https://doi.org/10.1126/science.1256785}
  {\path{doi:10.1126/science.1256785}}.

\bibitem{Duer:2018sby}
M.~Duer, et~al., {Probing high-momentum protons and neutrons in neutron-rich
  nuclei}, Nature 560~(7720) (2018) 617--621.
\newblock \href {https://doi.org/10.1038/s41586-018-0400-z}
  {\path{doi:10.1038/s41586-018-0400-z}}.

\bibitem{Duer:2018sxh}
M.~Duer, et~al., {Direct Observation of Proton-Neutron Short-Range Correlation
  Dominance in Heavy Nuclei}, Phys. Rev. Lett. 122~(17) (2019) 172502.
\newblock \href {http://arxiv.org/abs/1810.05343} {\path{arXiv:1810.05343}},
  \href {https://doi.org/10.1103/PhysRevLett.122.172502}
  {\path{doi:10.1103/PhysRevLett.122.172502}}.

\bibitem{Schmookler:2019nvf}
B.~Schmookler, et~al., {Modified structure of protons and neutrons in
  correlated pairs}, Nature 566~(7744) (2019) 354--358.
\newblock \href {http://arxiv.org/abs/2004.12065} {\path{arXiv:2004.12065}},
  \href {https://doi.org/10.1038/s41586-019-0925-9}
  {\path{doi:10.1038/s41586-019-0925-9}}.

\bibitem{Cruz-Torres:2020uke}
R.~Cruz-Torres, et~al., {Probing Few-Body Nuclear Dynamics via $^3$H and $^3$He
  ($e,e'p$)pn Cross-Section Measurements}, Phys. Rev. Lett. 124~(21) (2020)
  212501.
\newblock \href {http://arxiv.org/abs/2001.07230} {\path{arXiv:2001.07230}},
  \href {https://doi.org/10.1103/PhysRevLett.124.212501}
  {\path{doi:10.1103/PhysRevLett.124.212501}}.

\bibitem{Schmidt:2020kcl}
A.~Schmidt, et~al., {Probing the core of the strong nuclear interaction},
  Nature 578~(7796) (2020) 540--544.
\newblock \href {http://arxiv.org/abs/2004.11221} {\path{arXiv:2004.11221}},
  \href {https://doi.org/10.1038/s41586-020-2021-6}
  {\path{doi:10.1038/s41586-020-2021-6}}.

\bibitem{CLAS:2020rue}
I.~Korover, et~al., {12C(e,e'pN) measurements of short range correlations in
  the tensor-to-scalar interaction transition region}, Phys. Lett. B 820 (2021)
  136523.
\newblock \href {http://arxiv.org/abs/2004.07304} {\path{arXiv:2004.07304}},
  \href {https://doi.org/10.1016/j.physletb.2021.136523}
  {\path{doi:10.1016/j.physletb.2021.136523}}.

\bibitem{Furnstahl:2001xq}
R.~Furnstahl, H.~Hammer, {Are occupation numbers observable?}, Phys. Lett. B
  531 (2002) 203--208.
\newblock \href {http://arxiv.org/abs/nucl-th/0108069}
  {\path{arXiv:nucl-th/0108069}}, \href
  {https://doi.org/10.1016/S0370-2693(01)01504-0}
  {\path{doi:10.1016/S0370-2693(01)01504-0}}.

\bibitem{Bogner:2012zm}
S.~Bogner, D.~Roscher, {High-momentum tails from low-momentum effective
  theories}, Phys. Rev. C 86 (2012) 064304.
\newblock \href {http://arxiv.org/abs/1208.1734} {\path{arXiv:1208.1734}},
  \href {https://doi.org/10.1103/PhysRevC.86.064304}
  {\path{doi:10.1103/PhysRevC.86.064304}}.

\bibitem{Anderson:2010aq}
E.~R. Anderson, S.~K. Bogner, R.~J. Furnstahl, R.~J. Perry, {Operator Evolution
  via the Similarity Renormalization Group I: The Deuteron}, Phys. Rev. C 82
  (2010) 054001.
\newblock \href {http://arxiv.org/abs/1008.1569} {\path{arXiv:1008.1569}},
  \href {https://doi.org/10.1103/PhysRevC.82.054001}
  {\path{doi:10.1103/PhysRevC.82.054001}}.

\bibitem{Tropiano:2020zwb}
A.~J. Tropiano, S.~K. Bogner, R.~J. Furnstahl, {Operator evolution from the
  similarity renormalization group and the Magnus expansion}, Phys. Rev. C
  102~(3) (2020) 034005.
\newblock \href {http://arxiv.org/abs/2006.11186} {\path{arXiv:2006.11186}},
  \href {https://doi.org/10.1103/PhysRevC.102.034005}
  {\path{doi:10.1103/PhysRevC.102.034005}}.

\bibitem{Brueckner:1955zzd}
K.~Brueckner, R.~Eden, N.~Francis, {High-Energy Reactions and the Evidence for
  Correlations in the Nuclear Ground-State Wave Function}, Phys. Rev. 98 (1955)
  1445--1455.
\newblock \href {https://doi.org/10.1103/PhysRev.98.1445}
  {\path{doi:10.1103/PhysRev.98.1445}}.

\bibitem{Pandharipande:1979bv}
V.~R. Pandharipande, R.~B. Wiringa, {Variations on a theme of nuclear matter},
  Rev. Mod. Phys. 51 (1979) 821--859.
\newblock \href {https://doi.org/10.1103/RevModPhys.51.821}
  {\path{doi:10.1103/RevModPhys.51.821}}.

\bibitem{Morales:2002qi}
J.~Morales, V.~R. Pandharipande, D.~G. Ravenhall, {Improved variational
  calculations of nucleon matter}, Phys. Rev. C 66 (2002) 054308.
\newblock \href {https://doi.org/10.1103/PhysRevC.66.054308}
  {\path{doi:10.1103/PhysRevC.66.054308}}.

\bibitem{Neff:2015xda}
T.~Neff, H.~Feldmeier, W.~Horiuchi, {Short-range correlations in nuclei with
  similarity renormalization group transformations}, Phys. Rev. C 92~(2) (2015)
  024003.
\newblock \href {http://arxiv.org/abs/1506.02237} {\path{arXiv:1506.02237}},
  \href {https://doi.org/10.1103/PhysRevC.92.024003}
  {\path{doi:10.1103/PhysRevC.92.024003}}.

\bibitem{Weiss:2023laq}
R.~Weiss, S.~Gandolfi, {Nuclear three-body short-range correlations in
  coordinate space}, Phys. Rev. C 108~(2) (2023) L021301.
\newblock \href {http://arxiv.org/abs/2301.09605} {\path{arXiv:2301.09605}},
  \href {https://doi.org/10.1103/PhysRevC.108.L021301}
  {\path{doi:10.1103/PhysRevC.108.L021301}}.

\bibitem{Pieper:2002ne}
S.~C. Pieper, K.~Varga, R.~B. Wiringa, {Quantum Monte Carlo calculations of
  A=9, A=10 nuclei}, Phys. Rev. C 66 (2002) 044310.
\newblock \href {http://arxiv.org/abs/nucl-th/0206061}
  {\path{arXiv:nucl-th/0206061}}, \href
  {https://doi.org/10.1103/PhysRevC.66.044310}
  {\path{doi:10.1103/PhysRevC.66.044310}}.

\bibitem{Metropolis:1953}
N.~Metropolis, A.~W. Rosenbluth, M.~N. Rosenbluth, A.~H. Teller, E.~Teller,
  \href{https://doi.org/10.1063/1.1699114}{Equation of state calculations by
  fast computing machines}, The Journal of Chemical Physics 21~(6) (1953)
  1087--1092.
\newblock \href {http://arxiv.org/abs/https://doi.org/10.1063/1.1699114}
  {\path{arXiv:https://doi.org/10.1063/1.1699114}}, \href
  {https://doi.org/10.1063/1.1699114} {\path{doi:10.1063/1.1699114}}.
\newline\urlprefix\url{https://doi.org/10.1063/1.1699114}

\bibitem{Hastings:1970}
W.~K. Hastings, \href{https://doi.org/10.1093/biomet/57.1.97}{{Monte Carlo
  sampling methods using Markov chains and their applications}}, Biometrika
  57~(1) (1970) 97--109.
\newblock \href
  {http://arxiv.org/abs/https://academic.oup.com/biomet/article-pdf/57/1/97/23940249/57-1-97.pdf}
  {\path{arXiv:https://academic.oup.com/biomet/article-pdf/57/1/97/23940249/57-1-97.pdf}},
  \href {https://doi.org/10.1093/biomet/57.1.97}
  {\path{doi:10.1093/biomet/57.1.97}}.
\newline\urlprefix\url{https://doi.org/10.1093/biomet/57.1.97}

\bibitem{Wiringa:2013ala}
R.~Wiringa, R.~Schiavilla, S.~C. Pieper, J.~Carlson, {Nucleon and nucleon-pair
  momentum distributions in $A \le 12$ nuclei}, Phys. Rev. C 89~(2) (2014)
  024305.
\newblock \href {http://arxiv.org/abs/1309.3794} {\path{arXiv:1309.3794}},
  \href {https://doi.org/10.1103/PhysRevC.89.024305}
  {\path{doi:10.1103/PhysRevC.89.024305}}.

\bibitem{Wiringa:single_distributions}
R.~Wiringa, Single-nucleon momentum distributions,
  \href{https://www.phy.anl.gov/theory/research/momenta/}{https://www.phy.anl.gov/theory/research/momenta/}
  (last updated: June 26, 2023).

\bibitem{Pudliner:1995wk}
B.~S. Pudliner, V.~R. Pandharipande, J.~Carlson, R.~B. Wiringa, {Quantum Monte
  Carlo calculations of A \ensuremath{<}= 6 nuclei}, Phys. Rev. Lett. 74 (1995)
  4396--4399.
\newblock \href {http://arxiv.org/abs/nucl-th/9502031}
  {\path{arXiv:nucl-th/9502031}}, \href
  {https://doi.org/10.1103/PhysRevLett.74.4396}
  {\path{doi:10.1103/PhysRevLett.74.4396}}.

\bibitem{Dudek:1982zz}
J.~Dudek, Z.~Szymanski, T.~R. Werner, A.~Faessler, C.~Lima, {Description of the
  high spin states in Gd-146 using the optimized Woods-Saxon potential}, Phys.
  Rev. C 26 (1982) 1712--1718.
\newblock \href {https://doi.org/10.1103/PhysRevC.26.1712}
  {\path{doi:10.1103/PhysRevC.26.1712}}.

\bibitem{Schwierz:2007ve}
N.~Schwierz, I.~Wiedenhover, A.~Volya, {Parameterization of the Woods-Saxon
  Potential for Shell-Model Calculations} (9 2007).
\newblock \href {http://arxiv.org/abs/0709.3525} {\path{arXiv:0709.3525}}.

\bibitem{Volya:woods_saxon}
A.~Volya, Nucracker, \href{http://www.volya.net}{http://www.volya.net}.

\bibitem{Lapikas:1993aa}
L.~Lapik{\'a}s,
  \href{https://www.sciencedirect.com/science/article/pii/037594749390630G}{Quasi-elastic
  electron scattering off nuclei}, Nuclear Physics A 553 (1993) 297--308.
\newblock \href {https://doi.org/https://doi.org/10.1016/0375-9474(93)90630-G}
  {\path{doi:https://doi.org/10.1016/0375-9474(93)90630-G}}.
\newline\urlprefix\url{https://www.sciencedirect.com/science/article/pii/037594749390630G}

\bibitem{Massella:2018xdj}
P.~Massella, F.~Barranco, D.~Lonardoni, A.~Lovato, F.~Pederiva, E.~Vigezzi,
  {Exact restoration of Galilei invariance in density functional calculations
  with quantum Monte Carlo}, J. Phys. G 47~(3) (2020) 035105.
\newblock \href {http://arxiv.org/abs/1808.00518} {\path{arXiv:1808.00518}},
  \href {https://doi.org/10.1088/1361-6471/ab588c}
  {\path{doi:10.1088/1361-6471/ab588c}}.

\bibitem{Chabanat:1997un}
E.~Chabanat, P.~Bonche, P.~Haensel, J.~Meyer, R.~Schaeffer, {A Skyrme
  parametrization from subnuclear to neutron star densities. 2. Nuclei far from
  stablities}, Nucl. Phys. A 635 (1998) 231--256, [Erratum: Nucl.Phys.A 643,
  441--441 (1998)].
\newblock \href {https://doi.org/10.1016/S0375-9474(98)00180-8}
  {\path{doi:10.1016/S0375-9474(98)00180-8}}.

\bibitem{Bennaceur:2005mx}
K.~Bennaceur, J.~Dobaczewski, {Coordinate-space solution of the
  Skyrme-Hartree-Fock-Bogolyubov equations within spherical symmetry. The
  Program HFBRAD (v1.00)}, Comput. Phys. Commun. 168 (2005) 96--122.
\newblock \href {http://arxiv.org/abs/nucl-th/0501002}
  {\path{arXiv:nucl-th/0501002}}, \href
  {https://doi.org/10.1016/j.cpc.2005.02.002}
  {\path{doi:10.1016/j.cpc.2005.02.002}}.

\bibitem{Tropiano:2022jjj}
A.~J. Tropiano, S.~K. Bogner, R.~J. Furnstahl, M.~A. Hisham, {Quasi-deuteron
  model at low renormalization group resolution}, Phys. Rev. C 106~(2) (2022)
  024324.
\newblock \href {http://arxiv.org/abs/2205.06711} {\path{arXiv:2205.06711}},
  \href {https://doi.org/10.1103/PhysRevC.106.024324}
  {\path{doi:10.1103/PhysRevC.106.024324}}.

\bibitem{Machleidt:2000ge}
R.~Machleidt, {The High precision, charge dependent Bonn nucleon-nucleon
  potential (CD-Bonn)}, Phys. Rev. C 63 (2001) 024001.
\newblock \href {http://arxiv.org/abs/nucl-th/0006014}
  {\path{arXiv:nucl-th/0006014}}, \href
  {https://doi.org/10.1103/PhysRevC.63.024001}
  {\path{doi:10.1103/PhysRevC.63.024001}}.

\bibitem{Entem:2017gor}
D.~R. Entem, R.~Machleidt, Y.~Nosyk, {High-quality two-nucleon potentials up to
  fifth order of the chiral expansion}, Phys. Rev. C 96~(2) (2017) 024004.
\newblock \href {http://arxiv.org/abs/1703.05454} {\path{arXiv:1703.05454}},
  \href {https://doi.org/10.1103/PhysRevC.96.024004}
  {\path{doi:10.1103/PhysRevC.96.024004}}.

\bibitem{Reinert:2017usi}
P.~Reinert, H.~Krebs, E.~Epelbaum, {Semilocal momentum-space regularized chiral
  two-nucleon potentials up to fifth order}, Eur. Phys. J. A 54~(5) (2018) 86.
\newblock \href {http://arxiv.org/abs/1711.08821} {\path{arXiv:1711.08821}},
  \href {https://doi.org/10.1140/epja/i2018-12516-4}
  {\path{doi:10.1140/epja/i2018-12516-4}}.

\bibitem{Hisham:2022jzt}
M.~A. Hisham, R.~J. Furnstahl, A.~J. Tropiano, {Renormalization group evolution
  of optical potentials: Explorations using a
  \textquotedblleft{}toy\textquotedblright{} model}, Phys. Rev. C 106~(2)
  (2022) 024616.
\newblock \href {http://arxiv.org/abs/2206.04809} {\path{arXiv:2206.04809}},
  \href {https://doi.org/10.1103/PhysRevC.106.024616}
  {\path{doi:10.1103/PhysRevC.106.024616}}.

\bibitem{Weiss:2021rig}
R.~Weiss, P.~Soriano, A.~Lovato, J.~Menendez, R.~B. Wiringa, {Neutrinoless
  double-\ensuremath{\beta} decay: Combining quantum Monte Carlo and the
  nuclear shell model with the generalized contact formalism}, Phys. Rev. C
  106~(6) (2022) 065501.
\newblock \href {http://arxiv.org/abs/2112.08146} {\path{arXiv:2112.08146}},
  \href {https://doi.org/10.1103/PhysRevC.106.065501}
  {\path{doi:10.1103/PhysRevC.106.065501}}.

\end{thebibliography}

\end{document}